# A Deterministic TCP Bandwidth Sharing Model

Wolfram Lautenschlaeger
Bell Labs
Alcatel-Lucent
Stuttgart, Germany

***Abstract*—Traditionally TCP bandwidth sharing has been investigated mainly by stochastic approaches due to its seemingly chaotic nature. Even though of great generality, the theories deal mainly with expectation values, which is prone to misinterpretation with respect to the Quality-of-Experience (QoE). We disassemble TCP operating conditions into dominating scenarios and show that bandwidth sharing alone follows mostly deterministic rules. From the analysis we derive significant root causes of well-known TCP aspects like unequal sharing, burstiness of losses, global synchronization, and on buffer sizing. We base our model on a detailed analysis of bandwidth sharing experiments with subsequent mathematical reproduction.**

## I. Introduction

The Transmission Control Protocol (TCP) is used for reliable data transmission over best effort packet networks, i.e. networks, where packets randomly get lost without further notification. It has been standardized by the IETF, starting mainly with RFC 793 up to recently RFC 6937. The particular strength of TCP is its end-to-end flow control to cope with the enormous variety of network conditions along the transmission path. This way it laid the basis for the majority of Internet applications over the past 25 years. The large variety of impacting factors, however, makes it nearly impossible to give a closed form analysis of the TCP performance.

Most papers investigate the topic in great detail but still rely on expectation values rather than on actual traffic values. The resulting macroscopic performance figures are average values that give little insight on how to treat and improve the underlying real time process. For example adaptive video delivery over an IP network (IPTV) assumes that competing video streams share a common bottleneck in a way where all users get nearly the same video quality (i.e. bit rate). Indeed, most TCP theories state that (under certain circumstances) the expected bit rates for all bottleneck sharing flows are equal. However, as we will show later, the actual TCP bit rate of a particular (video) stream could deviate from its fair share by a factor of up to three for an extended period. The corresponding quality of experience (QoE) degradation cannot be detected by inspecting expectation values.

From source to destination a TCP flow typically traverses several links with quite different conditions. It could (a) fill a link alone (e.g. the end user's access link); (b) it could jointly, with few others fill a link, competing for a fair share of the link capacity (e.g. a broadband access link with several active home users), and (c) it could traverse a network with uncounted cross traffic flows, altogether creating a certain random packet loss and delay jitter. A flow cannot be assigned solely to one of the conditions. It might start in a server (capacity sharing (b)), pass the Internet (cross traffic (c)), and finally fill a home access link (single flow (a)). But most of the time, only one of them effectively limits the flow bit rate. An access limited flow (a) cannot compete for a fair share on other links since it is rate limited, anyway. Vice versa, a competition limited flow (b) might not be able to fill a subsequent access link even if it is alone there. The same holds for flows that do not sufficiently ramp up due to the general loss rate in the network (c).

In this paper we are mostly concerned with capacity sharing (b) of a limited number of competing flows. Obviously it includes the single flow condition (a) as a special case. We will show how capacity sharing synchronizes among flows to keep up the 100% link utilization. Furthermore we show how it creates a random loss process for the particular flow. The derived distribution of flow bit rates in response to the random loss process leads to the general cross traffic condition (c) with an independent random loss process.

We start our investigation with capacity sharing experiments, identify characteristic features of the traces, and then formulate analytic expressions that reproduce the characteristic features simply and concisely. With this approach we get a deep understanding of the bandwidth sharing process down to the timescale of one round trip time. We show that bandwidth sharing follows almost deterministic rules with only one exception – the random mapping of packet drops to the particular flows.

From the analysis we derive best case buffer size requirements that cannot be undercut in any case and we identify the root cause of global synchronization effects that generate worst case buffer demand and correspondingly large latency and queuing jitter. We do not need any a-priory postulate of traffic burstiness, but we show that the sharing process by itself generates an exceptional bursty loss process.

The paper is structured as follows: In section II we introduce our traffic model. Section III describes the experimental framework, data conditioning, and identification of relevant traffic features. In section IV we do a mathematical step by step analysis of the identified traffic features. Section V gives an overview of related work. Finally section VI summarizes our results.

## II. THE TRAFFIC MODEL

### A. Basic TCP Terms

To accomplish its mission, TCP chops the data into segments, sends the segments over the packet network to the destination, reassembles the data, and sends acknowledgements of correctly received data back from the receiver to the transmitter. This enables the identification of lost segments, which is then signaled back to the transmitter. The transmitter retransmits until all data has reached the destination. The segment size is limited by the underlying packet network technology. For Ethernet the *maximum segment size* (*MSS*) is typically in the range of 12000 bit (1.5kB).

The acknowledgments arrive one *round trip time* (*RTT*) later than the original segments have been sent out. Thus, for a greedy source there exist at any time a number of segments that have been sent out but that have not been acknowledged yet, which are commonly called "segments in flight". Van Jacobsen [2] introduced the *congestion window* (*cwnd*) as dynamically controlled limitation of the "segments in flight" for adaptation of the transmission data rate to the actual network conditions.

Acknowledgements are cumulative by their nature (i.e. "all data received so far are correct"). RFC 1122 [3] recommends one acknowledgment per two received full segments, which is commonly called as "delayed ACK". We define the *acknowledgement ratio a*, here $a$=2, as the average number of received (maximum) segments per returned acknowledgement.

### B. Congestion Avoidance

The TCP transmitter probes the available transmission capacity by adjusting its congestion window *cwnd* according to following rules:

1. In case of congestion experienced (packet loss detected): Halve the congestion window
   $cwnd \leftarrow cwnd/2$
2. In any other case: With each arriving acknowledgement increase the congestion window by 1/*cwnd.* This rule results in an almost linear increase of the congestion window with every elapsed round trip time, i.e. $cwnd \leftarrow cwnd+1$ if every segment would be acknowledged. Due to the delayed ACK, however:
   $cwnd \leftarrow cwnd+1/a$

Intuitively, those two adaptation rules result in a saw tooth like oscillation around an optimum that makes best use of the actually available transmission capacity. The procedure belongs to a class of additive increase, multiplicative decrease congestion avoidance algorithms (AIMD, [4]).

We assume sufficiently high-performance network conditions with low loss rates, selective acknowledgements, and fast recovery. Therefore we ignore the negligible extra traffic load due to retransmission and accordingly the window handling during the recovery phase. Furthermore we disregard slow start, time-out recovery, and application or advertised receive window limited operation.

### C. Sub-RTT dynamics

In this section we elaborate a subtle but crucial difference between bandwidth limitation by a capacity bottleneck and bandwidth limitation by congestion. The differentiation is indispensable for the conditioning of the experimental data of section III.C.

Given a certain congestion window, the delivery of packets exhibits the so called self-clocking effect, Fig. 1: A transmitter (Tx) might initially send packets at much too high rate until its congestion window is exhausted. Nevertheless the packets cannot cross the network faster than the narrowest intermediate link permits (bottleneck). Excess packets need to wait in a buffer before the bottleneck link. Downstream, the inter-packet distances of the bottleneck link are preserved, not only on subsequent higher capacity links, but also in the corresponding reverse acknowledgment chain. After an initial round trip the transmitter is removing acknowledged packets form its congestion window at the same pace as the acknowledgements arrive. This leads to further segments released exactly at a distance that fits to the intermediate bottleneck. As a result, self-clocking creates a smooth end-to-end packet flow at the bottleneck's capacity limit.

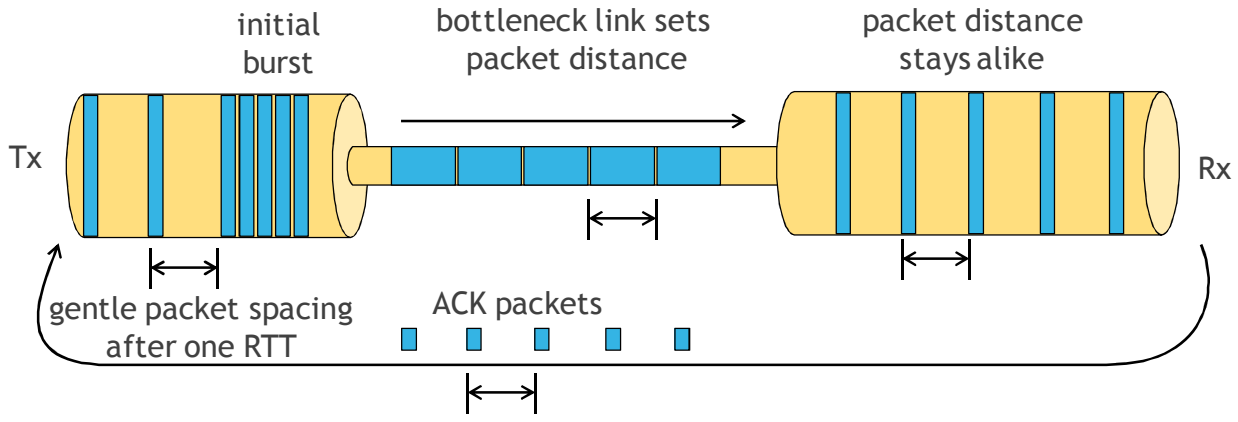

Fig. 1. Adaptation to bottleneck capacity by self-clocking

In case of bandwidth sharing the link capacity is not a bottleneck for a single flow. Other flows might constrict the available bandwidth, so that a single flow experiences macroscopically the equivalent of a bottleneck. On smaller time scales, however, there exists a remarkable difference: Self clocking, as in a hard bottleneck, does not necessarily occur, as illustrated in Fig. 2. Bursts of packets from different flows traverse sequentially the shared link. They are preserved on the reverse acknowledgement chain, and finally reproduce themselves in the next round.

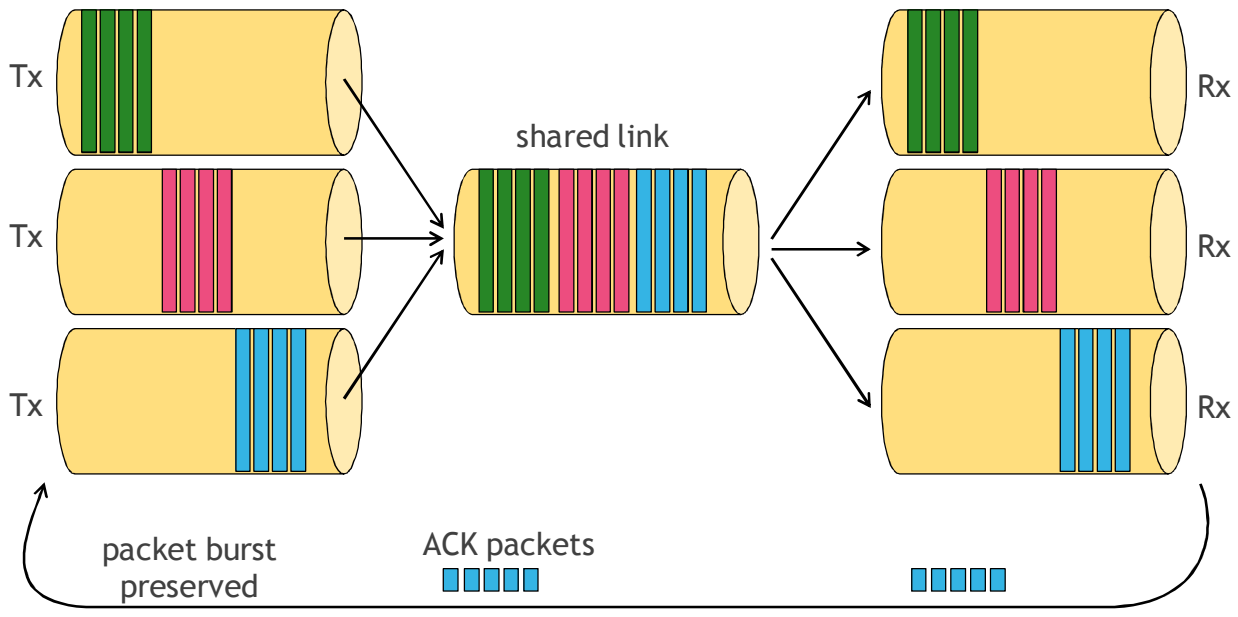

Fig. 2. Burst conservation in case of bandwidth sharing (no self-clocking)

The effect is well known [5], [6]. It can be easily reproduced in experiment. Real networks, however, due to the variety of impacting factors, operate certainly somewhere in between of self-clocking and burst conservation. It is important to note that for time scales shorter than the round trip time, link sharing flows are not as flat as self-clocking might suggest.

## III. Experiments

### A. Test Network

The test network, Fig. 3, comprised two state-of-the-art server machines and a carrier grade service router, all equipped with 10G Ethernet line cards. Operating system of both servers was Ubuntu 12.04 LTS, Linux Kernel 3.2. Most network parameter settings were left at their default values, except TCP buffer size in order to avoid congestion window limitations. All processor offload mechanisms of the Ethernet driver (lro, gro, tso, gso, etc.) were switched off. Instead of CUBIC, TCP Reno was enabled as congestion avoidance algorithm. A configurable number of pairs of greedy TCP transmitters and receivers were run on the server machines. The bottleneck link was a rate limited router output port with a buffer in front, worth of around 30ms queuing delay at the actual port rate. With bottleneck rates below 1Gbit/s no congestion could occur in the line cards or network stacks on the end systems. We patched the TCP stack of the receiving machine with a round trip time emulation, which implements an adjustable constant delay of the acknowledgments at sub-millisecond precision.

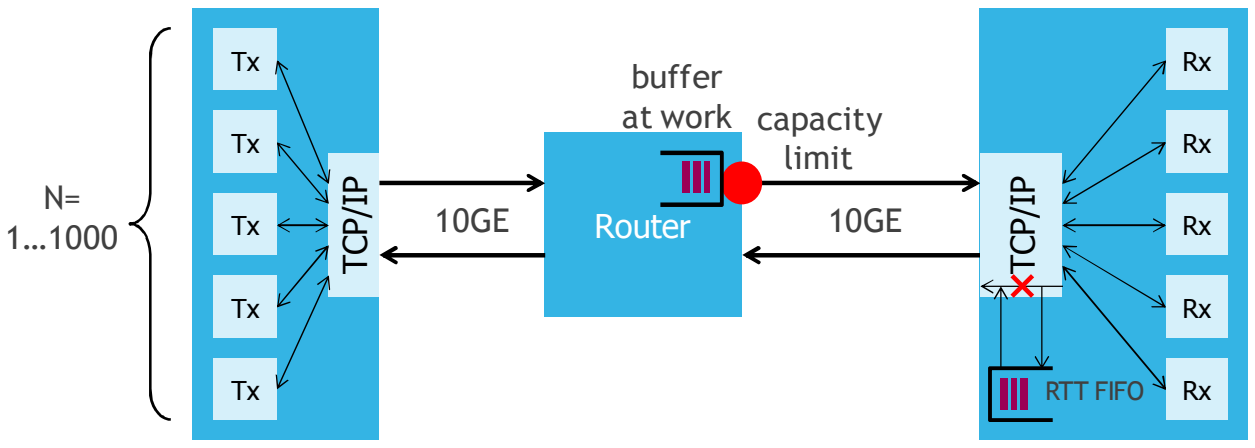

Fig. 3. Experimental set-up

### B. Raw Data

In the following we analyze an experiment with $N$=10 concurrent TCP flows, a bottleneck transmission capacity of 100Mbit/s, and a round trip time of $RTT$=100ms. Fig. 4 shows traces of the outgoing traffic over a period of 50 seconds, sampled[1] in 100 millisecond intervals. Please note that the traces do not include any startup phase. Packet capturing was started several minutes after flow initialization for making sure that all flows left the TCP slow start period and reached steady state congestion avoidance operation.

The total bit rate of all flows combined is almost flat at the capacity limit of the bottleneck link, with only a few dips. The per-flow bit rate traces exhibit an envelope with the expected saw tooth structure of section II.B. But within the envelope, the flow bit rates show arbitrary dips scattered at a seconds time scale. The saw tooth structure of flows raises the question, how they could form the flat total sum, even before traversing the queue. Obviously the scattering below the envelope cannot be ignored. But altogether it is not obvious by which means the scattered flow bit rates are coordinated.

[1] The reference to a sampling interval is inevitable. There is no meaningful interpretation of such term like instantaneous bit rate of packet traffic.

### C. Conditioning of Captured Data

Most publications on TCP experiments do not give an explanation for this behavior. Further analysis is typically based on expectation values. We identified as a cause of the burstiness of flow rates the unsteady round trip times due to queue variations in the router. Fig. 5 shows the effective round trip time evolution of the experiment in Fig. 4 as derived from an acknowledgement analysis in the captured packet sequence. As consequence, the sub-RTT packet bursts of Fig. 2 circulate at a period that reasonably differs from the 100ms sampling rate, which causes aliasing artifacts[2] leading to the scattering as shown in Fig. 4.

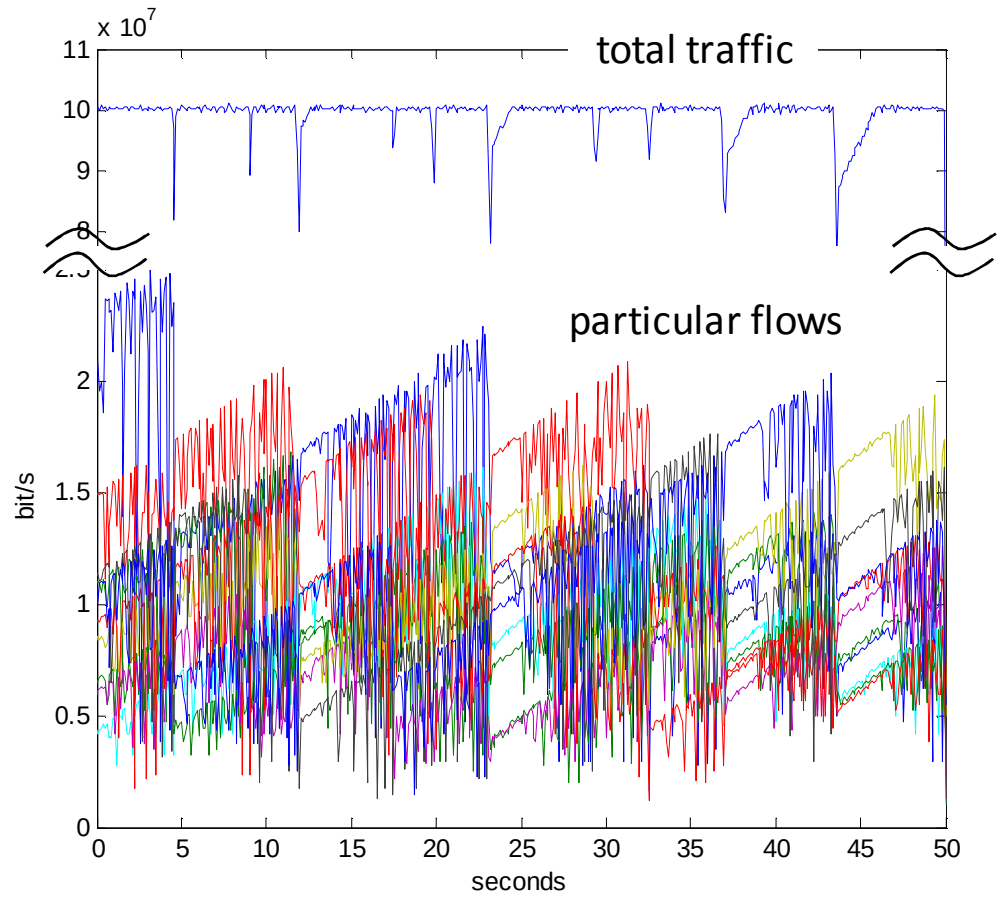

Fig. 4. Experimental data, 10 flows, 100ms RTT, sampled at 100ms intervals

Low pass filtering (e.g. larger intervals or sliding window averages) to mitigate aliasing artifacts cannot be used here, since the process characteristics itself (saw tooth) would be blurred[3]. Instead, we chose to resample the trace to a sequence of variable sized, nonoverlapping intervals, where the particular interval duration corresponds to the round trip time *at the given moment*. Formally: Given a sequence of packet arrivals ($\tau_j$, $x_j$) (arrival time and size) and a round trip time function $RTT(t)$ of Fig. 5 then the resampled bit rate trace ($t_i$, $b_i$) is:

$$t_{i+1} = t_i + RTT(t_i), \quad b_i = \frac{1}{RTT(t_i)} \sum_{j:t_{i-1}<\tau_j \le t_i} x_j \qquad (1)(2)$$

Fig. 6 is based on the identical raw data as Fig. 4 but with remarkably better insight thanks to the resampling technique. The burstiness at seconds time scale disappeared. Now the flow trajectories are made of smooth sections surrounded by well-defined abrupt changes. Even if not perfect, it is intuitively clear that the total sum of these traces could be flat, indeed.

We identified the following details for further mathematical analysis: Step down as a direct consequence of congestion window halving, step up of other flows due to queue relaxation

[2] Aliasing is a well-known signal distortion effect in digital signal processing
[3] Essentially, the use of expectation values is an infinite low pass filtering.

(section IV.A), sections of convergence towards fair share (from above and below, sections IV.B and C), and sections of linear increase (pure saw tooth) due to empty queue. The congestion incidents are, as expected, linked to the queue maxima of Fig. 5, but they are in no way linked to the actual flow rates. The consequences are investigated in section IV.D. Finally, if we zoom into the details, one can detect that at least 2 and up to 6 flows participate in any congestion event, i.e. simultaneously reduce their congestion window (global synchronization, section IV.E).

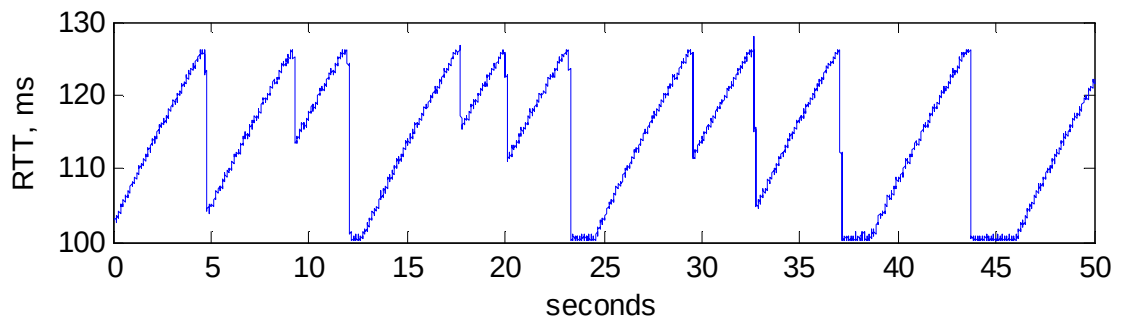

Fig. 5. Effective round trip time as seen by TCP

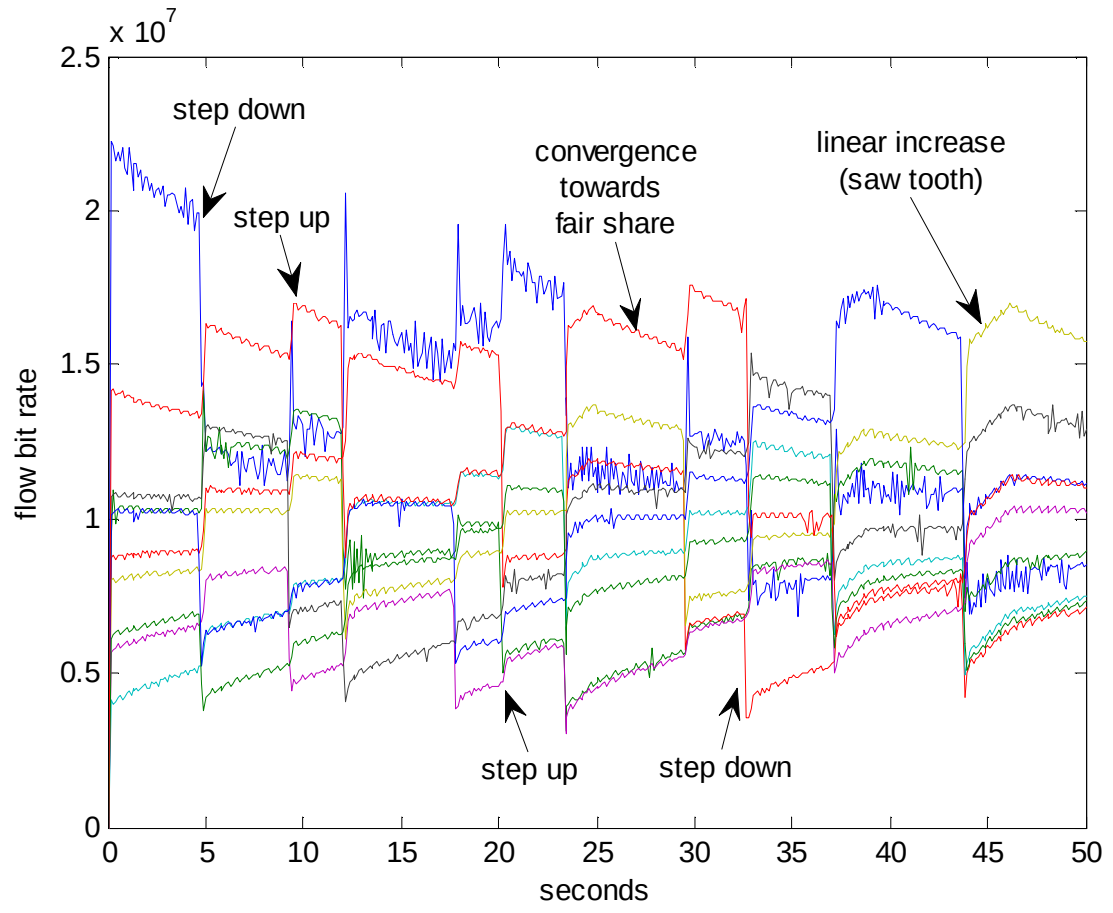

Fig. 6. Experimental data, resampled to effective *RTT* intervals

## IV. Analysis

### A. Step up, step down

The congested link has a capacity $C$ in bit/s to carry traffic. In case of $N$ concurrent flows, a calculative fair share of each flow would be $b_s=C/N$. The actual flow bit rate $b$, however, might be slightly different. It is the amount of data "in flight" (the congestion window *cwnd* multiplied by packet size *MSS*) divided by the round trip time *RTT*.

$$b = \frac{cwnd \cdot MSS}{RTT} \tag{3}$$

When a flow is hit by a packet loss, then the transmitter reduces its congestion window *cwnd* and correspondingly its sending rate by half. $b_{hit}$ denotes here the actual bitrate of the one flow that was hit by a packet loss, and at the moment, it was actually hit.

$$\Delta b_{down} = b_{hit}/2 \tag{4}$$

$$\Delta cwnd = \frac{\Delta b_{down} RTT}{MSS} = \frac{b_{hit}}{2}\frac{RTT}{MSS} \tag{5}$$

If the queue in front of the congested link is not empty at that moment, it continues to drain at rate $C$, but it is filled at reduced input bit rate. Correspondingly, its size goes down by an amount of $\Delta Q_{down}$ until the input bit rate catches up with the drain rate (or the queue goes empty). This increase in input bit rate has nothing to do with a congestion window increase. It is mainly due to the reduced queuing delay:

$$\Delta RTT = \Delta Q / C \tag{6}$$

If the queue does not drain to zero, the total sum of all flows equals the link capacity *before* and *after* the queue reduction.

$$\left(\sum b_i\right)_{before} = \left(\sum b_i\right)_{after} = C \tag{7}$$

Taking into account Eq. 3, we see that the sum of all congestion windows (of all participating transmitters) is proportional to the actual round trip time.

$$\frac{\left(\sum cwnd\right)_{before}}{RTT_{before}} = \frac{\left(\sum cwnd\right)_{after}}{RTT_{after}} = \frac{C}{MSS} \tag{8}$$

Accordingly, assuming that the congestion window reduction of Eq. 5 is the only change in $\sum cwnd_i$, the change in round trip time equals:

$$\Delta RTT_{down} = \frac{MSS}{C}\Delta cwnd = \frac{b_{hit}}{2}\frac{RTT}{C} \tag{9}$$

For a first estimation we assume that all flows oscillate around their fair share at $b_s=C/N$. In this ideal scenario, the switching point is at $b_{hit}=4/3b_s$ and after halving the congestion window at $2/3b_s$, resulting in $b=b_s$ in average. Thus

$$\Delta RTT_{down} \cong \frac{2b_s}{3}\frac{RTT}{C} = \frac{2}{3}\frac{RTT}{N}. \tag{10}$$

The queuing delay reduction $\Delta RTT_{down}$, together with Eq. 6, corresponds to a queue size reduction of

$$\Delta Q_{down} = \Delta RTT_{down} \cdot C = \frac{b_{hit}}{2} RTT. \tag{11}$$

With the simplifications above we get

$$\Delta Q_{down} \cong \frac{2}{3}\frac{C \cdot RTT}{N}. \tag{12}$$

**Conclusion:** Bandwidth delay product divided by the number of sharing flows is the absolute best case buffer dimensioning. Below that limit, 100% TCP utilization of a bottleneck link cannot be reached. This holds for all TCP flavors that use *cwnd* halving in response to packet loss. Larger buffers might be required due to: (a) The hit flow might deviate from the assumption of fair share by a factor of 3 (section IV.D). (b) Multiple flows could experience packet losses almost simultaneously, thus multiplying the required buffer size (global synchronization, section IV.E). (c) Issues, like random packet multiplex set their own lower bounds for buffer sizing, unrelated to TCP, in range of 80-100 packets. We refer here to the standard queuing theory, e.g. [9].

Returning to the per flow analysis, we see that, due to the reduced round trip time, any flow of actual rate $b$ experiences an acceleration $\Delta b_{up}$:

$$\Delta b_{up} = b_{after} - b_{before} = \frac{cwnd \cdot MSS}{RTT_{after}} - \frac{cwnd \cdot MSS}{RTT_{before}} \quad (13)$$

$$\Delta b_{up} = cwnd \cdot MSS \frac{\Delta RTT_{down}}{RTT_{before} RTT_{after}} = b \frac{\Delta RTT_{down}}{RTT_{after}} \quad (14)$$

Together with Eq. 9, where $RTT$ is actually $RTT_{before}$, and $RTT_{after}=RTT_{before}-\Delta RTT_{down}$, we finally get:

$$\Delta b_{up} = b \frac{b_{hit}}{2C - b_{hit}} \quad (15)$$

Taking into account that the hit flow experiences first a halving of its rate, but afterwards participates with its remaining (halved) bit rate in the step up, we get:

$$\Delta b = \begin{cases} \frac{b_{hit}}{2}\left(\frac{b_{hit}}{2C - b_{hit}} - 1\right) & \text{if this flow was hit} \\ b \cdot \frac{b_{hit}}{2C - b_{hit}} & \text{if other flow was hit} \end{cases} \quad (16)$$

The general Equation 16 becomes clearer if we look at two extreme values: (1) In the single flow case we get $b_{hit}=C$, which results in $\Delta b=0$, i.e. no steps at all. The flow is filling the capacity at its limit all the time. (2) In case of many flows (N>>1), together with the simplifications, yielding to Eq. 10, we get

$$\Delta b \cong \begin{cases} -\frac{2b_s}{3} & \text{if this flow was hit} \\ \frac{2b_s}{3N} & \text{if other flow was hit} \end{cases} \quad (17)$$

Obviously, the sum of up steps by not affected flows compensates the single down step by the flow that was hit by a packet loss, so that the total sum remains almost flat as shown in Fig. 4.

By replacing $b_{hit}/2$ in Eq. 11 with an arbitrary change in the offered bit rate $\Delta B$ we get

$$\Delta Q = RTT \cdot \Delta B \cdot \quad (18)$$

**Conclusion:** Aggregated TCP traffic is able to absorb an offered load change of $\Delta B$ without losing throughput and without packet loss if the queue has room for $RTT \cdot \Delta B$ (up and down, respectively). The adaptation happens in just one round trip time. This could be the analytical basis for unsteady state bandwidth sharing scenarios.

*B. Queue evolution between steps*

In times between packet losses, all flows increase their congestion windows according to the congestion avoidance algorithm, section II.B. With $N$ flows and an acknowledgment ratio $a$, the sum of all congestion windows *before* and *after* a single RTT interval (not to be mixed up with *before* and *after* *cwnd* reduction steps of section IV.A) increases according to

$$\left(\sum cwnd\right)_{after} = \left(\sum cwnd\right)_{before} + N/a \cdot \quad (19)$$

By similar considerations as of Eq. 8 we get

$$\frac{\Delta cwnd}{\Delta RTT} = \frac{C}{MSS}, \text{ with } \Delta cwnd = N/a, \text{ or} \quad (20)(21)$$

$$\Delta RTT = \frac{MSS}{C}\frac{N}{a}, \quad (22)$$

which is the change of $RTT$ in a time interval $\Delta t=RTT(t)$. The corresponding differential equation and its solution, starting at $RTT(0)=RTT_0$, are

$$\frac{\Delta RTT}{\Delta t} = \frac{MSS \cdot N}{C \cdot a}\frac{1}{RTT(t)}, \quad (23)$$

$$RTT(t) = \sqrt{2\frac{MSS \cdot N}{C \cdot a} \cdot t + RTT_0^2} \cdot \quad (24)$$

The linearization of Eq. 24 in the neighborhood of $RTT_0$ is

$$RTT(t) \cong RTT_0 + \frac{MSS \cdot N}{C \cdot a \cdot RTT_0} \cdot t \quad (25)$$

Fig. 7 (left) shows a comparison of the analytical solutions in Eq. 24 and 25 with the experimental trace of Fig. 5.

The queue size evolution can be estimated by taking into account Eq. 6 and starting from an empty queue at $t=0$:

$$Q(t) \cong \Delta RTT \cdot C = \left(RTT(t) - RTT_0\right) \cdot C \quad (26)$$

$$Q(t) \cong \frac{MSS \cdot N}{a \cdot RTT_0} \cdot t \quad (27)$$

Now we estimate the mean time interval $T$ between packet losses by equating the queue step down of Eq. 12 and the linearized increase of Eq. 27.

$$\frac{2}{3}\frac{RTT\cdot C}{N}=\frac{MSS\cdot N}{a\cdot RTT_0}T \qquad (28)$$

$$T=\frac{2a}{3}\frac{RTT^2\cdot C}{MSS\cdot N^2} \qquad (29)$$

A word of caution is necessary here: $T$ is a synthetic average parameter. In practice, the loss process is extremely bursty due to global synchronization, section IV.E. Long intervals of no losses at all are followed by dense series of losses. Anyway, since loss events are equally distributed among flows, the average loss interval per flow $T_s$ can be calculated as follows ($T_s= N\cdot T$):

$$T_s=\frac{2a}{3}\frac{RTT^2}{MSS}b_s \qquad (30)$$

In our example the loss interval per flow $T_s$ is 11 seconds.

The reciprocal value of $T$ is the packet loss *rate*. Divided by the total packet rate, we get the packet loss probability $P_{loss}$.

$$P_{loss}=\frac{1}{T}\Big/\frac{C}{MSS}=\frac{3}{2a}\left(\frac{MSS\cdot N}{RTT\cdot C}\right)^2 \qquad (31)$$

By solving the above for the link capacity we get the well know formulas for macroscopic TCP performance [7], [8] at aggregate and at flow level.

$$C=\sqrt{\frac{3}{2a}}\frac{MSS}{\sqrt{P_{loss}}\,RTT}N \qquad (32)$$

$$b_s=\sqrt{\frac{3}{2a}}\frac{MSS}{\sqrt{P_{loss}}\,RTT} \qquad (33)$$

Taking into account Eq. 3, we get for the equilibrium congestion window (flow is acquiring its fair share):

$$cwnd_s=\sqrt{\frac{3}{2a}}\frac{1}{\sqrt{P_{loss}}} \qquad (34)$$

*C. Flow evolution between steps*

The bit rate of a flow, starting at arbitrary value $b_0$, changes by two factors: (a) Increase of the congestion window by $1/a$, and (b) increase of the round trip time according to Eq. 24. Keeping in mind Eq. 3 we get a bit rate change within one round trip time of

$$\Delta b=\frac{(cwnd+1/a)MSS}{RTT+\Delta RTT}-\frac{cwnd\cdot MSS}{RTT}, \qquad (35)$$

$$\Delta RTT=RTT(RTT)-RTT(0)=\frac{MSS\cdot N}{C\cdot a}, \qquad (36)$$

which results in

$$\Delta b=\frac{MSS}{a}\frac{(1-b/b_s)}{RTT+\Delta RTT}. \qquad (37)$$

Assuming $\Delta RTT<<RTT$, we get

$$\Delta b=\frac{MSS}{a\cdot RTT}\left(1-\frac{b}{b_s}\right). \qquad (38)$$

The corresponding time interval is $\Delta t=RTT(t)$, hence

$$\frac{\Delta b(t)}{\Delta t}=\frac{MSS\cdot(b_s-b(t))}{a\cdot b_s\cdot RTT(t)^2} \qquad (39)$$

It is obvious from Eq. 39 that the flow bit rate converges from an arbitrary starting point $b_0$ towards the fair share $b_s$ as expected from Fig. 6. The solution, incorporating $RTT(t)$ of Eq. 24, is as follows:

$$b(t)=\frac{b_0-b_s}{\sqrt{\frac{2MSS}{a\cdot b_s RTT_0^2}t+1}}+b_s \qquad (40)$$

Fig. 7 (right) shows the coincidence (within the bit-rate uncertainty) of the analytical solution of Eq. 40. with two arbitrary sections of the experimental data.

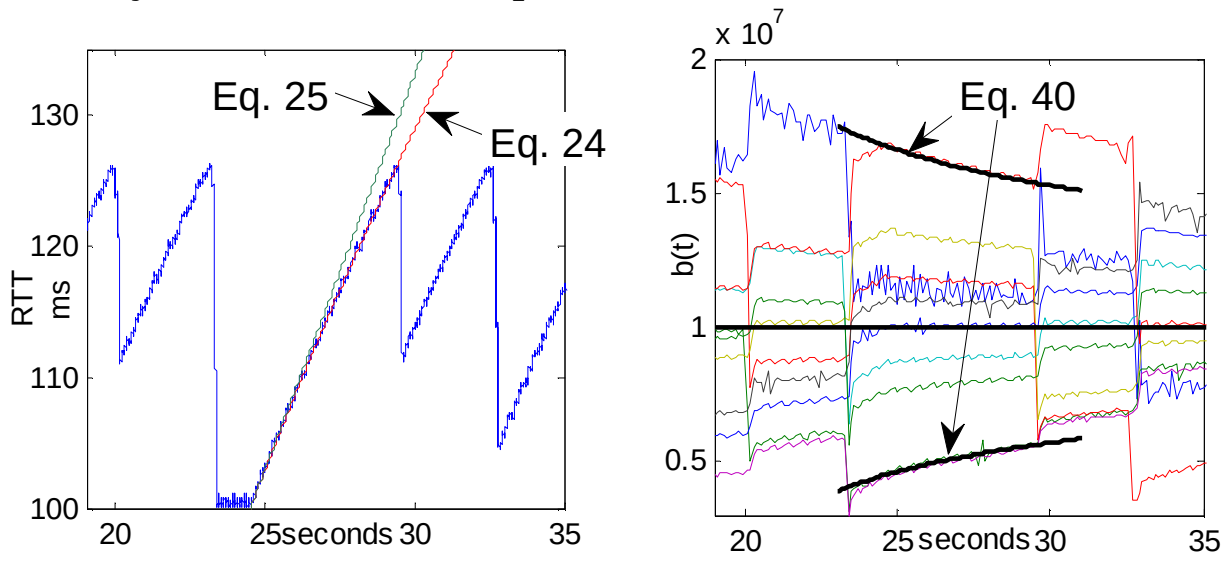

Fig. 7. Comparison experimental data vs. analytical expression ($MSS$=12000bit, $C$=100Mbit/s, $N$=10, $a$=2, $RTT_0$=0.1s, $t_0$=24.5s)

**Conclusion:** Between loss events the queue undergoes a monotonic increase, while the flows are converging towards their fair share bit rate. There is no flow related extremum triggering the subsequent loss. The mapping of losses to flows is random.

*D. Random drop mapping to flows*

The calculations above or a careful inspection of Fig. 6 reveals that bit rate steps down due to packet drops might affect any of the sharing flows, from biggest to smallest. There is no fairness at all on that. That's why we had to leave open the initial flow bit rates for the subsequent processes ($b_{hit}$ and $b$ of Eq. 16; $b_0$ of Eq. 40). Since the drop mapping to flows is random, the actual flow bit rate is random as well, even though the reaction to drop events and the evolution between drops is almost deterministic.

The probability distribution of the flow bit rate can be obtained by investigating the evolution of the congestion window as a continuous Markov chain. (We stick here to a method from [9].) Fig. 8 shows a fragment of the Markov chain, where the state nodes correspond to the actual congestion window size *cwnd*, and transition arcs correspond to transition rates in a small time interval $\Delta t$. The upper part shows the additive increment, which is solely dependent on the round trip time *RTT* and the acknowledgment ratio *a*. The lower part shows the *cwnd* halving in response to packet losses. Here, the transition rate depends on the packet loss probability $P_{loss}$. Multiplied by the actual packet rate (*cwnd*/*RTT*), we get the packet loss *rate* (lost packets per second). Further multiplied by the time interval $\Delta t$, we get the transition rate. This accounts for the fact that, even though the drop probability $P_{loss}$ is equal for all flows, the hit *rate* of a particular flow depends on the amount of packets sent, so that larger flows are more likely affected than smaller ones.

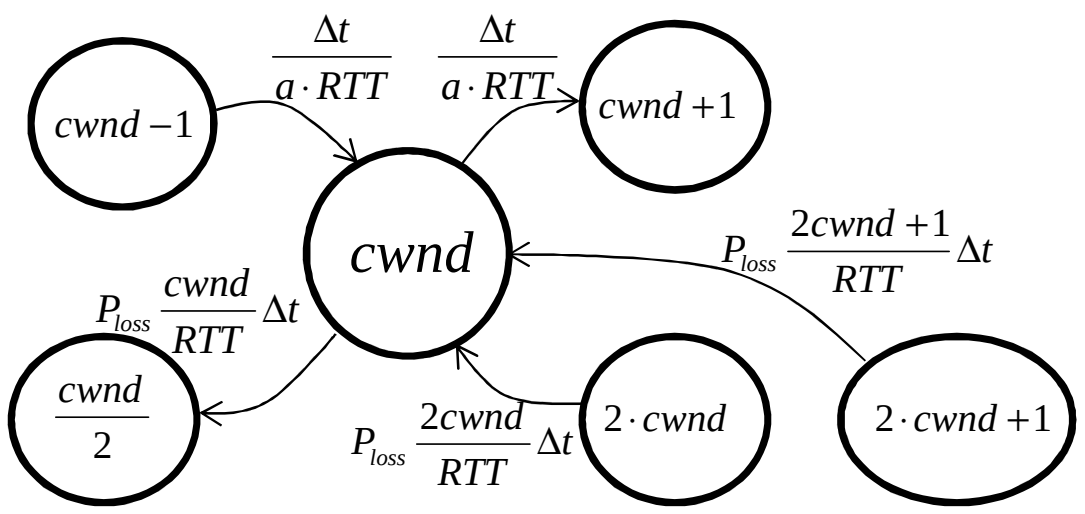

Fig. 8. Fragment of the congestion window state diagram

The state probabilities $p_i$ of *cwnd* to be in state $i \in [1, cwnd_{\max}]$ form a set of linear equations. The so called node equations are built with the assumption that in steady state each node should be reached at the same rate, as it is left. Otherwise the node probability would not be in equilibrium.

$$(1/a)\, p_{i-1} + 2iP_{loss} \cdot p_{2i} + (2i+1)P_{loss} \cdot p_{2i+1} = (P_{loss} i + 1/a)\, p_i \tag{41}$$

In matrix notation the corresponding state probability vector $P_{cwnd} = \left(p_1, p_2, \cdots, p_{cwnd_{\max}}\right)^T$ fulfills following equilibrium equation:

$$P_{cwnd} = A \cdot P_{cwnd} \tag{42}$$

With the shortcut $P=a \cdot P_{loss}$ the transition matrix *A* (with e.g. $cwnd_{max}$=7) looks like this:

$$A = \begin{bmatrix} 0 & \frac{2P}{1} & \frac{3P}{1} & 0 & 0 & 0 & 0 \\ \frac{1}{1+2P} & 0 & 0 & \frac{4P}{1+2P} & \frac{5P}{1+2P} & 0 & 0 \\ 0 & \frac{1}{1+3P} & 0 & 0 & 0 & \frac{6P}{1+3P} & \frac{7P}{1+3P} \\ 0 & 0 & \frac{1}{1+4P} & 0 & 0 & 0 & 0 \\ 0 & 0 & 0 & \frac{1}{1+5P} & 0 & 0 & 0 \\ 0 & 0 & 0 & 0 & \frac{1}{1+6P} & 0 & 0 \\ 0 & 0 & 0 & 0 & 0 & \frac{1}{7P} & 0 \end{bmatrix} \tag{43}$$

The state probability vector $P_{cwnd}$ is an eigenvector of the matrix *A* corresponding to eigenvalue 1. Alternatively, a numerical solution can be obtained by replacing one of the component equations by the normalizing condition $\Sigma\ p_i = 1$. Then, the bit rate distribution is the state probability vector, scaled according to Eq. (3). Fig. 9 shows the numerically evaluated probability distribution of flow bit rates with a parameter set conforming to the experiment of Fig. 6. An analog result has been published in [10]. It is obvious that the actual bit rate *b* of any of the sharing flows might deviate from its fair share value $b_s$ in a range of roughly $1/3b_s<b<3b_s$. As we are looking at the steady state probabilities, there is no further long term convergence towards fair share. Even flows that reached their fair share might subsequently go down to only 1/3 of that bit rate. Only the carried data volume, i.e. integral *b* over time *T*, converges eventually towards $b_s \cdot T$. We leave out for space restrictions an analysis of the convergence speed. But a first hint could be the loss interval per flow $T_s$ of Eq. 30. Intuitively, multiple of such intervals need to pass by for convergence of the carried data volume.

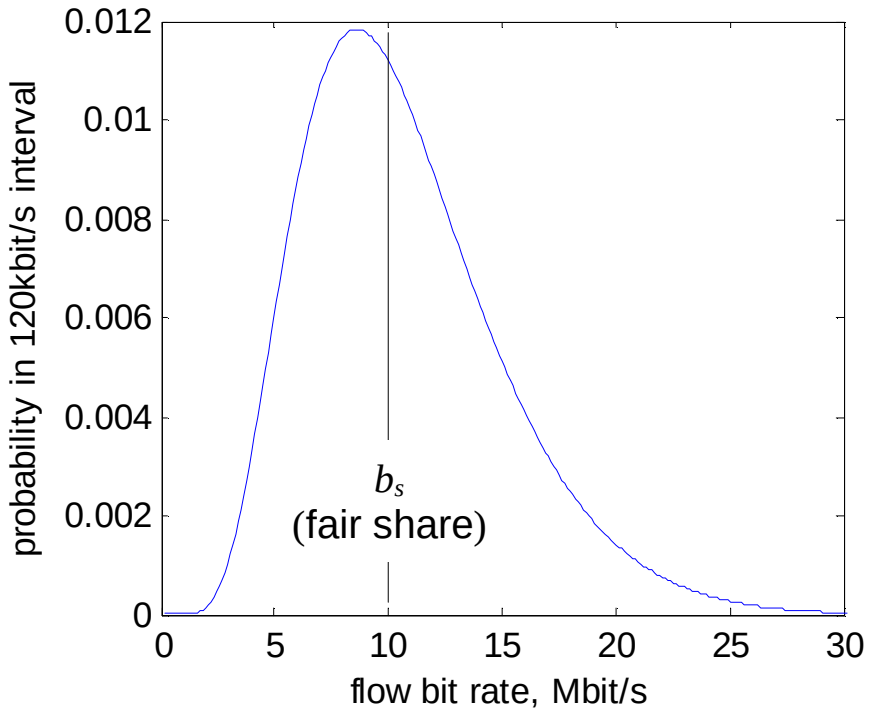

Fig. 9. Probability distribution of flow bit rates

The cumulative distribution function of *cwnd* can be empirically approximated by a log normal distribution with good fit in the center part, but large deviations in the far tails:

$$\mathrm{F}(cwnd) = \Phi\left[\frac{\ln(cwnd) - \mu}{\sigma\sqrt{2}}\right] \tag{44}$$

With $\Phi$ – the cumulative standard normal distribution function, $\sigma$ = 0.41=*const*., and $\mu=\ln(cwnd_s)$ – the logarithmic fair share congestion window of Eq. 34. The latter is indicating that our distribution function is in line with the macroscopic analysis of [7]. Fig. 10 shows a comparison of our numeric solution with a log normal approximation at different loss probabilities.

**Conclusion:** TCP flow bit rates might deviate from their expectation value at least by a factor of three for an extended period of time, longer than just one “saw tooth” interval. This holds for all scenarios except the single flow case.

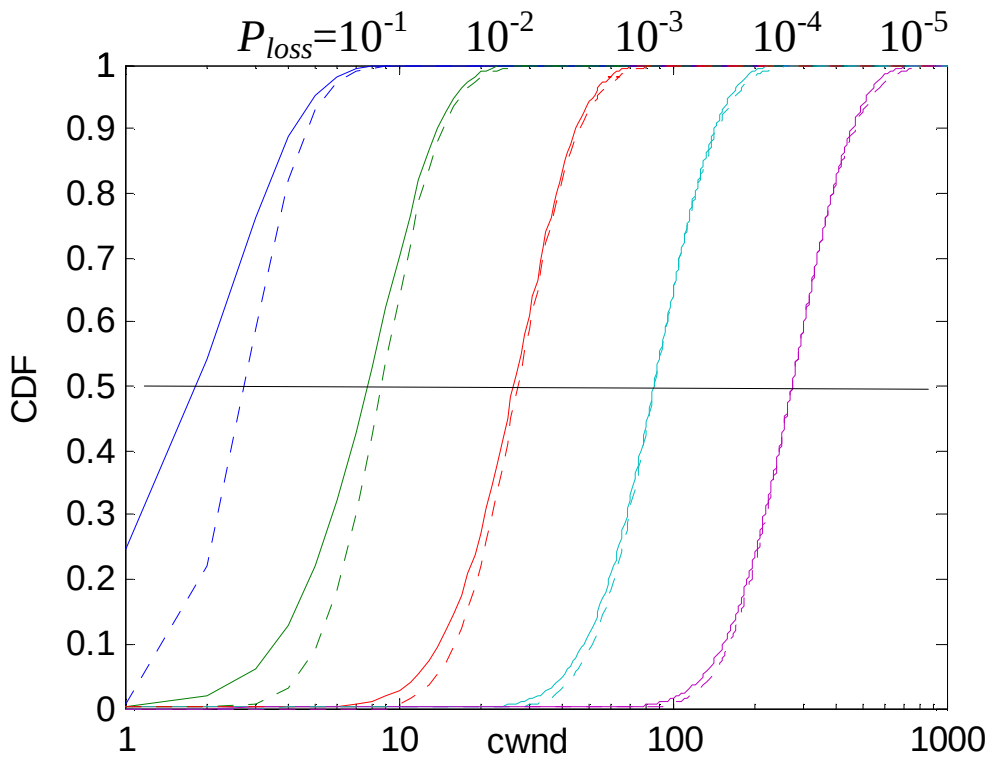

Fig. 10. Cumulative distribution function of the congestion window; a=2; dashed lines – log normal distribution

### E. Global synchronization

The step up, step down analysis of section IV.A suffers from the fact that typically more than one packet loss occurs in one congestion event, which causes more than one flow to reduce its bit rate by half. The effect is well known under the term global synchronization. As consequence, the bit rate changes according to Eq. 16, as well as the queue size reductions of Eq. 12 simply add up without any statistical averaging.

The reason behind global synchronization is hidden in the queue evolution right after the first packet drop. Up to this moment, the sum of all sharing flows was in equilibrium with the bottleneck capacity. After the first drop it takes *at least one round trip* until the bit rate reduction can take effect. During that interval, all transmitters continue to increase their congestion window. The sum of all *cwnd* increments during one *RTT* is $N/a$, cf. Eq. 19. Since at that moment the buffer is already full, these extra packets get lost and cause up to $N/a$ flows to reduce their bit rate by half. (In our initial experiment 5 flows of the 10.) In practice the number of affected flows might be lower, since some of them are hit repeatedly within the *RTT* interval, which they answer, however, only by one congestion window reduction (fast recovery [11]).

We verified the effect experimentally with up to $N$=110 flows and found typically 30-50% of the flows participating in a rate reduction during one round trip time. The quota was even higher with CUBIC due to the more aggressive *cwnd* increase.

A synchronization event is not to be understood as a solid back-to-back train of packet losses. It is rather a period of slightly increased loss probability. In our experiments, the largest congestion event contained 50 losses. But during that *RTT* interval 8300 other packets successfully crossed the queue. Hence, the loss ratio *during the burst* was as low as $6 \cdot 10^{-3}$.

Fig. 11 is a sample trace of retransmissions (at $N$=30). The distance between bursts is in range of 5 – 7 seconds, which fits to the mean drop distance $T_s$ of Eq. 30 (taking into account that half of the flows participates in one burst). Below, in the zoomed image of the burst (at 24.3 seconds), we see 11 separate retransmissions that are spread over a period of 120 milliseconds, i.e. the round trip forwarding plus queuing delay.

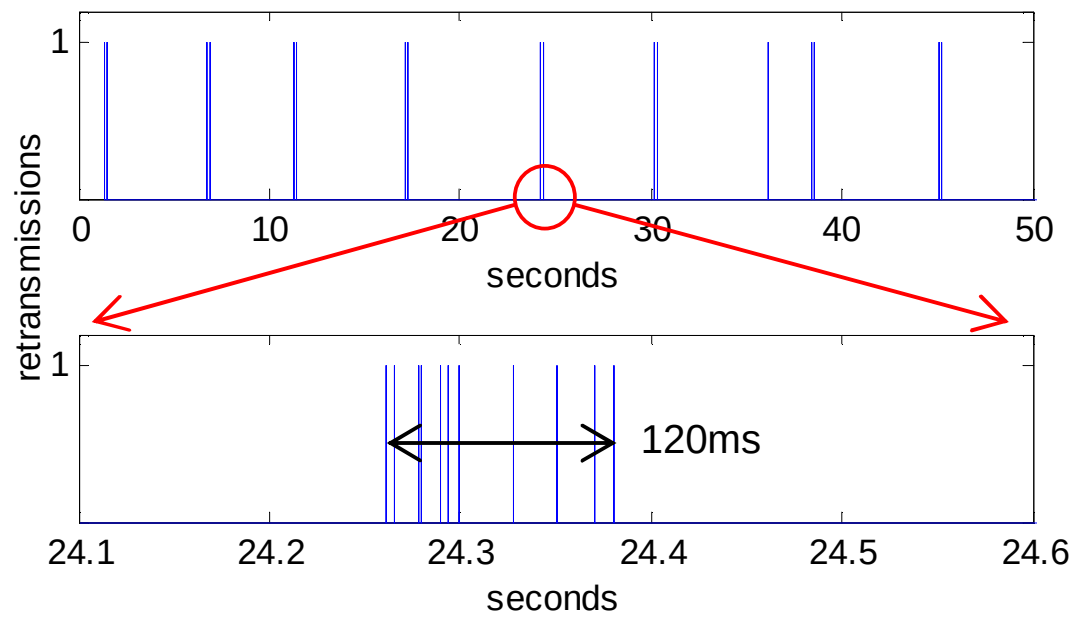

Fig. 11. Bursts of packet losses due to global synchronization

In practice, global synchronization is limited by two effects: (1) Large number of flows and (2) spreading of round trip times: When the raising edge of the saw tooth gets faster than one round trip time, then subsequent congestion events overlap, which essentially cuts their duration below one *RTT*.

With the global synchronization effect in mind, the buffer size dimensioning according to Eq. 12 needs to be revised. The queue size reduction of Eq. 12 could appear simultaneously $N/a$ times, which yields:

$$\Delta Q_{global} \cong \frac{2}{3a} C \cdot RTT \tag{45}$$

**Conclusion:** TCP traffic with global synchronization requires buffer according to the bandwidth delay product rule for 100% link utilization, independent of the number of participating flows. Active Queue Management (AQM) could reduce this demand by a factor of $N$, the number flows, *but not more*, cf. Eq. 12.

## V. Related Work

Mathis et al. [7] revealed the macroscopic TCP performance formula by idealized deterministic consideration. Padhye et al. [8] rebased the formula on a stochastic approach and extended it by the impact of the retransmission workload and time-outs, which is particularly relevant in high loss conditions. Damjanovic and Welzl [1] take further into account that in the bandwidth sharing scenario the observed loss bursts might trigger less congestion window reductions than isolated losses would do. Parvez, Mahanti, and Williamson [12] investigate the performance impact of the NewReno extensions [11] particularly in the presence of loss bursts. In both papers burstiness of losses is postulated and not derived from the sharing process itself. Bogoiavlenskaia [10] derived a numeric model of the steady state congestion window probability distribution. All the mentioned papers operate on expectation values, not on actual traffic values.

The additive increase multiplicative decrease (AIMD) class of congestion avoidance algorithms is known to be the only binary feedback algorithm that is fair, efficient, and stable at the same time [4]. TCP Reno [2] is a straight implementation of AIMD and serves as basis for almost all practically relevant TCP implementations. Recent Linux distributions enable by default a heuristic extension of the additive increase part,

CUBIC [13]. It reduces the *RTT* bias of TCP Reno by faster congestion window increase in case of large deviation from fair share and at larger round trip times. The faster increase rates result in a generally more robust but also more aggressive behavior with larger loss rates.

Active Queue Management (AQM) is used to control buffering. Industry standard is Random Early Detection (RED) [14]. It prevents global synchronization by randomizing the drops. Its only (but persistent) problem is the selection of appropriate parameter settings. The raising awareness of problems due to long standing queues [5] initiated recently new interest in AQM like e.g. CoDel [15], or PIE [16].

The absence of global synchronization on core network links with many flows and large round trip time spreading is generally accepted. With this assumption Appenzeller, Keslassy, McKeown [17] refined the bandwidth delay product rule by a factor $1/\sqrt{N}$. It negates global synchronization but accepts the uncoordinated loss assignment to flows. As such it lies at the geometric mean between our best (Eq. 12) and worst (Eq. 45) cases. Nevertheless, there might be cases, where the assumptions do not hold. E.g. the round trip times are not spread at all on interconnection lines between distant data centers. Frequent arrival and departure of flows (which is the rule in practice) put additional burden on the queue responsiveness (Eq. 18). Vu-Brugier et al. [18] raised serious concerns on this topic underpinned with field results.

Finally, the 100% link utilization target has been questioned more and more, particularly with the typically random traffic offer [5], [19]. Queues below the 100% link load cannot sustaining build up. The queuing delay remains negligibly low. This operating condition, also called *link transparency* or *flow-through* operation, is given in our sample trace at positions 24, 38, and 45 seconds from start. How to keep a shared link in transparent mode is discussed in [20]. It assumes TCP flows being shaped at the end-system access links, in opposite to the present paper with unlimited (greedy) traffic sources.

## VI. Conclusion

This paper is focused on transmission capacity sharing by greedy TCP flows. We show experimentally how aliasing artifacts render typical TCP flows into seemingly chaotic bursty processes. But we can show how they translate into piecewise deterministic functions that follow explicit mathematical formulas. While inspired by the experiments, the formulas are generic, not relying on any empirical parameter fitting.

From the mathematical analysis we derive bounds for buffer sizing. The lower bound at bandwidth delay product *divided by the number of flows* should not be undercut in any case, while the upper bound at the full bandwidth delay product is required, if the traffic experiences global synchronization among flows. Improvements by AQM are limited to the range between both bounds, which is particularly important for single- or few flow cases. The bounds are valid for all *cwnd* halving TCP flavors.

We show how capacity sharing generates a random loss process per flow. A numerically obtained probability distribution reveals potential flow bit rate deviations from expectation value by more than factor three for extended periods of time. This could severely impact the Quality of Experience (QoE) in applications like adaptive video streaming.

In our analysis we do not need any a-priori postulate on traffic burstiness. Moreover we show under which queuing conditions TCP traffic is able to absorb abrupt traffic offer changes without losing throughput and without extra losses. And we show how multiple sharing flows create bursts of packet losses without being bursty by themselves. By this way we explain the frequently observed effect of global synchronization. With the analysis of flow trajectory and probability distribution we show the mechanisms and to which extent TCP flow rates can be controlled from a routers perspective.

## Acknowledgment

This work has been funded in parts by the German Bundesministerimum für Bildung und Forschung (BMBF) in scope of project SASER under grant No. 16BP12200.

## References

[1] D. Damjanovic, M. Welzl, An extension of the TCP steady-state throughput equation for parallel flows and its application in MulTFRC, IEEE/ACM Trans. Networking, 19.6, 2011

[2] Van Jacobson, Congestion avoidance and control, Proc. SIGCOMM ´88, 1988

[3] Requirements for Internet Hosts -- Communication Layers, IETF, RFC 1122, 1989

[4] Dah-Ming Chiu, Raj Jain, Analysis of the increase and decrease algorithms for congestion avoidance in computer networks, J. Computer Networks and ISDN Systems, 17.1, 1989

[5] Van Jacobson, A rant on queues, A talk presented at MIT Lincoln Labs, Lexington, MA, 2006, online: http://www.pollere.net/Pdfdocs/QrantJul06.pdf

[6] Hao Jiang , Constantinos Dovrolis, Why is the internet traffic bursty in short time scales, Proc. SIGMETRICS ´05, 2005

[7] M. Mathis, J. Semke, J. Mahdavi, and T. Ott, The Macroscopic Behavior of the TCP Congestion Avoidance Algorithm, Comput. Commun. Rev., 27.3, 1997, pp. 67–82.

[8] J. Padhye, V. Firoiu, D. Towsley, and J. Kurose, “Modeling TCP throughput: A simple model and its empirical validation,” in Proc. ACM SIGCOMM, 1998, pp. 303–314.

[9] Handbook Teletraffic Engineering, ITU-D Study Group 2 Question 16/2, 2008.

[10] Bogoiavlenskaia, O., Markovian Model of Internetworking Flow Control, Kalashnikov Memorial Seminar, Petrozavodsk, Информационные процессы, 2.2, 2002

[11] S. Floyd, T. Henderson, and A. Gurtov, The NewReno modification to TCP’s fast recovery algorithm, IETF, RFC 3782, Apr. 2004.

[12] N. Parvez, A. Mahanti, and C. Williamson, An Analytic Throughput Model for TCP NewReno, IEEE/ACM Trans. Networking, 18.2, 2010

[13] Sangtae Ha, Injong Rhee, and Lisong Xu, CUBIC: A new TCP-friendly high-speed TCP variant, ACM SIGOPS Operating Systems Review, 42.5, 2008, pp. 64-74

[14] S. Floyd, Van Jacobsen, Random Early Detection Gateways for Congestion Avoidance, IEEE/ACM Trans. Networking, 1.4, 1993

[15] K. Nichols**,** Van Jacobson, "Controlling Queue Delay", ACM Queue – Networks, 2012

[16] R. Pan, et al., PIE: A lightweight control scheme to address the bufferbloat problem, IETF, draft-pan-tsvwg-pie-00, 2012

[17] G. Appenzeller, I. Keslassy, and N. McKeown, Sizing router buffers, Proc. ACM SIGCOMM ´04, 2004.

[18] G. Vu-Brugier, R. S. Stanojević, D. J. Leith, R. N. Shorten, A critique of recently proposed buffer-sizing strategies, ACM SIGCOMM Computer Communication Review, 37.1, 2007

[19] S. B. Fredj, T. Bonald, A. Proutiere, G. Régnié, J.W. Roberts, Statistical bandwidth sharing: a study of congestion at flow level, Proc. ACM SIGCOMM ´01, New York, 2001.

[20] W. Lautenschlaeger, F. Feller, Light-weight traffic parameter estimation for online bandwidth provisioning, 24th International Teletraffic Congress, ITC24, Kraków, 2012